# Modal Analysis of Power System with High CIG Penetration Based on Impedance Models

Le Zheng, Jiajie Zheng, Jiajian Lin, and Chongru Liu*

*Abstract*— This paper explores the modal analysis of power systems with high Converter-Interfaced Generation (CIG) penetration utilizing an impedance-based modeling approach. Traditional modal analysis based on the state-space model (MASS) requires comprehensive control structures and parameters of each system element, a challenging prerequisite as converters increasingly integrate into power systems and their internal specifics remain largely inaccessible. Conversely, the proposed modal analysis based on the impedance model (MAI) leverages only the impedance port characteristics to pinpoint system elements significantly influencing unstable modes. This study is the first to confirm the theoretical equivalency between MASS and MAI in terms of transfer functions, eigenvalues, and sensitivities, thus bridging the gap between detailed theoretical modeling and practical, accessible analyses. We further provide enhancements to the MAI method, including a revised element participation index, a transformer ratio-based admittance sensitivity adjustment, and an impedance splitting-based sensitivity analysis considering parameter variations. Validation through numerical simulations on a modified IEEE 14-bus system underscores the efficacy of our approach. By examining the interplay between different elements and system modes in high CIG environments, this study offers insights and a foundational framework for delineating the oscillatory modes' participation and stability characteristics of power systems with substantial CIG integration.

*Index Terms*— Modal Analysis, impedance model, Converter-Interfaced Generation, eigenvalue sensitivity.

## I. Introduction

With the increasing share of renewable energy in the global energy mix, the penetration of converter-interfaced generation (CIG) in power systems has increased significantly, which poses new stability challenges to power systems. Compared to the synchronous generators (SG), the dynamics of CIG exhibit wider and faster time scales in power angle, frequency, and voltage, potentially giving rise to distinct electromagnetic modes [1]. In the past decade, some oscillation events have been reported and related to the complex interactions among various resources and elements in power systems with high CIG penetration [2]. Moreover, the low inertial support characteristics of CIGs can cause the oscillations to propagate faster in the grid [3]. Consequently, developing new modeling and analyzing frameworks becomes imperative to address these evolving complexities [4].

Impedance modeling is one of the most effective tools for evaluating the small-signal stability of power systems with high CIG penetration [5]. An impedance network transformation method is proposed in [6] to simplify the complexity of generating network impedance matrices for stability analysis. Ref. [7] proposes a stability criterion for positive modal damping, but it was only applicable to the weakly damped oscillation modes. Ref. [8] establishes a method to embed the dynamics of the coordinate transformation into the impedance model, which enables the total impedance modeling of the whole system. Nonetheless, while impedance-based analyses can ascertain system stability, they fall short in pinpointing the most influential elements or factors within the system.

In contrast, modal analysis based on the state-space model (MASS) stands out as a pivotal method for identifying the key factors based on linear algebra. MASS describes the degree of influence of the state variables on the modes based on the participation factors [9]. Nevertheless, state-space modeling necessitates detailed models and specific parameters for each system element, making it unsuitable for CIGs analysis. Due to commercial confidentiality, converter manufacturers typically only disclose the impedance model that describes the port characteristics.

Several studies have investigated methods for utilizing impedance models to identify the critical factors contributing to oscillations. Ref. [10] introduces the Resonance Mode Analysis technique to identify the bus that has the highest participation. Eigenvalue sensitivity is derived and utilized to determine the pivotal elements of the network that influence some specific oscillatory modes [11], whereas the eigenvalue trajectory is employed in [12] to evaluate how controller parameters impact harmonic instability. Ref. [13] discusses a frequency-domain modal analysis approach, calculating oscillatory modes by locating the zeros of the determinant of the loop impedance matrix or nodal admittance matrix.

Recently, modal analysis based on the impedance model (MAI) has been developed to facilitate root-cause tracing without the need for internal control details of the power supply [14]-[15]. The impedance participation factor is defined and used to locate the power supply that contributes most to the system instability. Together with the parameter participation factor and element participation factor, this method establishes

The work is supported by the National Natural Science Foundation of China under Grant 52307095.

The authors (corresponding author, email: chongru.liu@ncepu.edu.cn) are with the State Key Lab of Alternate Electrical Power System with Renewable Energy Sources, North China Electric Power University, Beijing 102206, China.

three layers/degrees of transparency for root-cause analysis, enabling investigations into various depths of the system. To expand the sensitivity analysis beyond just the power supply to encompass all network elements, the admittance sensitivity factor, and parameter sensitivity factor have been developed [15], realizing participation analysis in the impedance model for the whole system.

MAI is important for small-signal stability analysis in power systems with high CIG penetration, which can be adapted for broader scenarios. For example, [16] extends it to AC-DC hybrid systems considering the effect of the interlinking converter. In addition, the linkage between MASS and MAI is drawn using the diagonal element of the state matrix [14]. However, establishing the equivalence of these two modal analysis methods requires further theoretical investigation. This deeper exploration would solidify understanding and ensure accurate application across different system analyses. The main contributions of this paper can be summarized as:

(1) Based on a brief review of the MASS and MAI, the equivalence of the two modal analysis methods has been rigorously analyzed for the first time, in terms of transfer function, eigenvalue, and sensitivity. The equivalence analysis provides a solid theoretical foundation for the in-depth understanding and further improvement of MAI.

(2) To make MAI more applicable, enhancement has been made from three distinct perspectives. A new total evaluation index is proposed to fix the potential misjudgment when computing the admittance sensitivity factor. The effect of the transformer branch on the whole-system admittance matrix is derived to further reduce estimation error. In addition, the sensitivity of the impedance to its internal parameter is replaced with the equivalent admittance sensitivity factor, which is much easier to obtain.

The paper is structured as follows: Section II gives preliminary knowledge on MASS and MAI methods. Discussions on the equivalency between the two modal analysis methods are presented in Section III. Section IV provides enhancement details and practical implementations. Section V performs simulation verification with a modified IEEE 14-bus system. The last section concludes the paper.

## II. MODAL ANALYSIS THEORY

### A. Modal Analysis Based on the State-Space Model

The linearized state-space equation of a power system can be expressed as follows:
$$\Delta \dot{x} = A\Delta x + B\Delta u \\ \Delta y = C\Delta x + D\Delta u \quad (1)$$

where $\Delta x$ is the state vector, $\Delta u$ is the input vector, and $\Delta y$ is the output vector. $A$, $B$, $C$, and $D$ are the state matrix, input matrix, output matrix, and feedforward matrix, respectively.

The diagonalization of the state matrix $A$ by the $\Delta x = \Phi \Delta z$ coordinate transformation yields a diagonal state matrix.
$$\Lambda = \Psi A \Phi = \mathrm{diag}(\lambda_1 \cdots \lambda_i \cdots \lambda_n) \quad (2)$$
where $\lambda_i$ is the $i$-th eigenvalue of matrix $A$, $\phi_i$ and $\psi_i$ in $\Phi = [\phi_1 \cdots \phi_i \cdots \phi_n]$ and $\Psi = [\psi_1^T \cdots \psi_i^T \cdots \psi_n^T]^T$ are the right eigenvector and left eigenvector of $\lambda_i$, respectively. We have $\Phi\Psi = I$, where $I$ is the unit matrix.

The sensitivity of $\lambda_i$ to the $k$-th row and $j$-th column element $a_{kj}$ of the matrix $A$ is:
$$\frac{\partial \lambda_i}{\partial a_{kj}} = \psi_{ik} \phi_{ji} \quad (3)$$

where $\phi_{ji}$ and $\psi_{ik}$ are the $j$-th element of the right eigenvector $\phi_i$ and the $k$-th element of the left eigenvector $\psi_i$, respectively.

System participation matrix $P = [P_1 \cdots P_i \cdots P_n]$ is as follows.
$$P_i = \begin{bmatrix} p_{1i} \\ \vdots \\ p_{ki} \\ \vdots \\ p_{ni} \end{bmatrix} = \begin{bmatrix} \phi_{1i}\psi_{i1} \\ \vdots \\ \phi_{ki}\psi_{ik} \\ \vdots \\ \phi_{ni}\psi_{in} \end{bmatrix} \quad (4)$$

where $p_{ki}$ denotes the relative participation of the $k$-th state variable in the $i$-th mode, which equals the sensitivity of $\lambda_i$ to the $k$-th diagonal element $a_{kk}$ of matrix $A$ [9].

### B. Whole-System Impedance Modeling

Consider the system shown in Fig. 1, $\Delta I_j = \begin{bmatrix} \Delta i_{jd} & \Delta i_{jq} \end{bmatrix}^T$ is the dq axis component of the injected current at bus $j$ and $\Delta U_j = \begin{bmatrix} \Delta u_{jd} & \Delta u_{jq} \end{bmatrix}^T$ is the dq axis component of the voltage change at bus $j$.

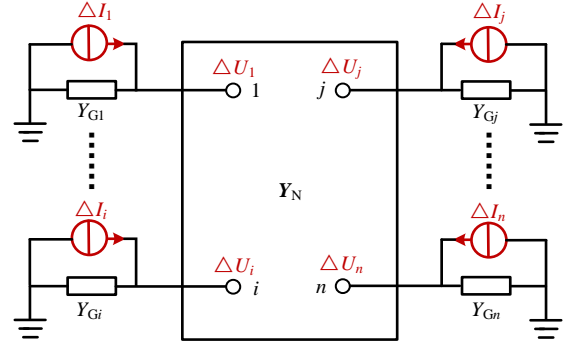

Fig. 1 System structure diagram.

$Y_{Gi}$ represents the admittance of the power supply connected to bus $i$ in the dq coordinate, which is a 2nd-order square matrix:
$$Y_{Gi} = Z_{Gi}^{-1} \quad (5)$$
where $Z_{Gi}$ denotes the impedance model of the power supply connected to bus $i$. If there is no power supply at bus $i$, there is $Y_{Gi} = 0$. The power supply admittance matrix of the system can be obtained from each impedance model as follows.
$$Y_G = \mathrm{diag}(Y_{G1} \cdots Y_{Gi} \cdots Y_{Gn}) \quad (6)$$
where $Y_G \in 2n \times 2n$ denotes the admittance matrix of the system power supply and $n$ is the number of system nodes. The impedance/admittance model of a power supply can be derived by state equations [5] as well as by measurements [18]. It is worth noting that in the impedance modeling of multi-machine power systems, the impedance model of each power supply derived through the state equation needs to be rotated by coordinates to ensure that the coordinate systems of all power supplies are aligned [8].

The node admittance matrix of the power network in Fig. 1 is

$$\mathbf{Y}_N = \begin{bmatrix} Y_{N11} & \cdots & Y_{N1j} & \cdots & Y_{N1n} \\ \vdots & & \vdots & & \vdots \\ Y_{Ni1} & \cdots & Y_{Nij} & \cdots & Y_{Nin} \\ \vdots & & \vdots & & \vdots \\ Y_{Nn1} & \cdots & Y_{Nnj} & \cdots & Y_{Nnn} \end{bmatrix} \quad (7)$$

where $Y_{Nij}$ denotes the mutual admittance between buses $i$ and $j$, and $Y_{Nii}$ denotes the self-admittance of bus $i$.

Taking the perturbation values of the injected current $\Delta I_i$ at each bus as input, and the corresponding voltage change $\Delta U_i$ as output:

$$\Delta u = [\Delta I_1 \cdots \Delta I_i \cdots \Delta I_n]^T \quad (8)$$

$$\Delta y = [\Delta U_1 \cdots \Delta U_i \cdots \Delta U_n]^T \quad (9)$$

The closed-loop transfer function matrix from input to output can be derived [8].

$$\Delta y = (\mathbf{I} + \mathbf{Z}_N \mathbf{Y}_G)^{-1} \mathbf{Z}_N \Delta u = \mathbf{Z} \Delta u \quad (10)$$

$$\mathbf{Z} = (\mathbf{I} + \mathbf{Z}_N \mathbf{Y}_G)^{-1} \mathbf{Z}_N = (\mathbf{Y}_G + \mathbf{Y}_N)^{-1} = \mathbf{Y}^{-1} \quad (11)$$

where $\mathbf{Z}_N = \mathbf{Y}_N^{-1}$ is the node impedance matrix of the network. $\mathbf{Z}$ and $\mathbf{Y}$ are called the whole-system impedance matrix and the whole-system admittance matrix, respectively. Thus, the power supply impedance/admittance characteristics are included in the whole-system impedance/admittance matrix.

*C. Modal Analysis Based on the Impedance Model*

The eigenvalue sensitivity of mode $\lambda$ with respect to an element with admittance $y$ is given in [15].

$$\frac{\partial \lambda}{\partial y} = \sum_{i=1}^n \sum_{j=1}^n (\frac{\partial \lambda}{\partial Y_{ij}} \cdot \frac{\partial Y_{ij}}{\partial y}) = -\mathrm{tr}(\mathrm{Res}_\lambda \mathbf{Z} \cdot \frac{\partial \mathbf{Y}}{\partial y}) \quad (12)$$

where $Y_{ij}$ denotes the mutual admittance between bus $i$ and bus $j$ in the whole-system admittance matrix $\mathbf{Y}$. tr() denotes the trace of the matrix. $\mathrm{Res}_\lambda \mathbf{Z}$ denotes the residue of the matrix $\mathbf{Z}$ at mode $\lambda$, which can be identified from the spectrum by rational approximation [17].

The expansion of the above equation yields:

$$\frac{\partial \lambda}{\partial y} = \begin{cases} -\mathrm{Res}_\lambda Z_{ii} & \text{Node } i \\ -\mathrm{Res}_\lambda (Z_{ii} + Z_{jj} - Z_{ij} - Z_{ji}) & \text{Branch } ij \end{cases} \quad (13)$$

Ref. [15] defines the admittance sensitivity factor of $y$ as:

$$s_{\lambda,y} = (\frac{\partial \lambda}{\partial y})^* \quad (14)$$

where $*$ denotes the matrix conjugate transpose. Then：

$$\Delta \lambda = \langle s_{\lambda,y}, \Delta y \rangle \quad (15)$$

where $\langle , \rangle$ denotes the Frobenius inner product, and the equation describes the amount of corresponding change in the mode caused by a change in the admittance of elements in the system.

To further consider the variation induced by the parameters $\rho$ to $y$, the parameter sensitivity factor is defined as:

$$s_{\lambda,\rho} = \langle s_{\lambda,y}, \frac{\partial y}{\partial \rho} \rangle \quad (16)$$

The amount of mode changes due to the variation of the parameter $\rho$ is:

$$\Delta \lambda = s_{\lambda,\rho} \cdot \Delta \rho \quad (17)$$

Thus, it is possible to perform modal analyses with different transparency for each element of the system using the three-layer chain rule [14]. Use Cauchy's inequality for (15):

$$|\Delta \lambda| = |\langle s_{\lambda,y}, \Delta y \rangle| \le \|s_{\lambda,y}\| \cdot \|\Delta y\| = \varepsilon \|s_{\lambda,y}\| \cdot \|y\| = |\Delta \lambda|_{\max} \quad (18)$$

where $\|\;\|$ denotes the Frobenius norm, and $\varepsilon$ denotes the proportion of change in $y$, which takes a generally small value. $\|s_{\lambda,y}\| \cdot \|y\|$ is defined as the first layer of MAI to measure the possible total participation degree of admittance $y$ in mode $\lambda$.

The second and third layers are:

$$\sigma_2 + j\omega_2 = \langle s_{\lambda,y}, y \rangle \quad (19)$$

$$\sigma_3 + j\omega_3 = \langle s_{\lambda,y}, \frac{\partial y}{\partial \rho} \rangle \quad (20)$$

The second layer gets the effect of the elements on the damping (real part) and the natural oscillation frequency (imaginary part) of that mode. In contrast to the second layer, in order to analyze the effect of specific parameters on the modes, the third layer needs information on the sensitivity of the admittance to the parameters.

III. EQUIVALENCE OF THE TWO MODAL ANALYSIS METHODS

In contrast to MASS, which requires detailed control structures and parameters to build the differential equations in state space, MAI can work with port characteristics and locate the elements that contribute most to the unstable modes. Although the two methods have different theoretical frameworks, they are essentially equivalent in terms of transfer function, eigenvalue, and sensitivity.

*A. Equivalence in terms of Transfer Functions*

The transfer function $G_{ss}$ derived using the state-space model (1) is:

$$\Delta y_{ss} = C(sI - A)^{-1} B \Delta u_{ss}$$
$$= C\Phi(sI - \Lambda)^{-1} \Psi B \Delta u_{ss} = G_{ss} \Delta u_{ss} \quad (21)$$

In contrast, the transfer function $G_i$ of Fig. 1 derived using the impedance model is shown in (22), which equals to the whole-system impedance matrix $\mathbf{Z}$.

$$\Delta y_i = \mathbf{Z} \Delta u_i = C_1(sI - A)^{-1} B_1 \Delta u_i = G_i \Delta u_i \quad (22)$$

This comes from the fact that the input/output of the impedance model is a subset of those in the state-space model. Consequently, the columns of $B_1$ are selected from the corresponding columns of $B$, and the rows of $C_1$ are those chosen from $C$. Take the system containing a single converter working in grid-forming (GFM) strategy connected to an infinity bus as an example. The input and output variables of

the state-space model are:
$$\Delta u_{ss} = [\Delta i_{d1}, \Delta i_{q1}, \Delta P, \Delta i_{d2}, \Delta i_{q2}]^T$$
$$\Delta y_{ss} = [\Delta u_{d1}, \Delta u_{q1}, \Delta \omega, \Delta \theta, \Delta u_{d2}, \Delta u_{q2}]^T \quad (23)$$

where $\Delta i_{d1}$, $\Delta i_{q1}$, and $\Delta P$ are the perturbation value of the dq-axis injection current of GFM and the input power on the DC side. $\Delta u_{d1}$, $\Delta u_{q1}$, $\Delta \omega$, and $\Delta \theta$ are the change values of the dq-axis terminal voltage, output frequency and angle of GFM. $\Delta i_{d2}$ and $\Delta i_{q2}$ are the perturbation value of dq-axis injection current of the infinity bus, whereas $\Delta u_{d2}$ and $\Delta u_{q2}$ are the perturbation values of the dq-axis voltage.

The input and output of the impedance model are shown in (24). $B_1$ is constructed by the 1st, 2nd, 4th, and 5th columns of $B$, and $C_1$ contains the 1st, 2nd, 5th, and 6th rows of $C$.
$$\Delta u_i = [\Delta i_{d1}, \Delta i_{q1}, \Delta i_{d2}, \Delta i_{q2}]^T$$
$$\Delta y_i = [\Delta u_{d1}, \Delta u_{q1}, \Delta u_{d2}, \Delta u_{q2}]^T \quad (24)$$

Therefore, in terms of transfer function expression, the two methods are equivalent.

*B. Equivalence in terms of Eigenvalues*

Apply the Laplace transform to the state transfer matrix $e^{At}$ of the state-space model:
$$L[e^{At}] = (sI - A)^{-1} \quad (25)$$

where $L[]$ denotes the Laplace transform. Since any eigenvalue $\lambda_i$ is a pole of $(sI - A)^{-1}$, substituting the Laplace operator $s$ with $\lambda_i$ leads to
$$|\lambda_i I - A| = 0 \quad (26)$$

$\lambda_i$ is also a pole of the whole-system impedance matrix [13]:
$$|0I - Y(\lambda_i)| = 0 \quad (27)$$

Calculating the eigenvector corresponding to $\lambda_i$ for $A$ is equivalent to determining that for the eigenvalue 0 for $Y(\lambda_i)$, which is called the critical resonance mode [11]. Thus, the two methods are equivalent in terms of eigenvalue analysis.

*C. Equivalence in terms of Sensitivity*

In state space, following (3) we can get:
$$\frac{\partial \lambda_i}{\partial A} = \begin{bmatrix} \frac{\partial \lambda_i}{\partial a_{11}} & \cdots & \frac{\partial \lambda_i}{\partial a_{1j}} & \cdots & \frac{\partial \lambda_i}{\partial a_{1n}} \\ \vdots & & \vdots & & \vdots \\ \frac{\partial \lambda_i}{\partial a_{i1}} & \cdots & \frac{\partial \lambda_i}{\partial a_{ij}} & \cdots & \frac{\partial \lambda_i}{\partial a_{in}} \\ \vdots & & \vdots & & \vdots \\ \frac{\partial \lambda_i}{\partial a_{n1}} & \cdots & \frac{\partial \lambda_i}{\partial a_{nj}} & \cdots & \frac{\partial \lambda_i}{\partial a_{nn}} \end{bmatrix}$$
$$= (\phi_i \otimes \psi_i)^T = (\phi_i \psi_i)^T \quad (28)$$

where $\otimes$ denotes the Kronecker product and T is the transpose of a matrix.

Taking the residue at mode $\lambda_i$ for matrix $(sI - A)^{-1}$:

$$\text{Res}_{\lambda_i}(sI-A)^{-1} = \lim_{s \to \lambda_i}((s-\lambda_i) \cdot \Phi(sI-\Lambda)^{-1}\Psi)$$
$$= \begin{bmatrix} \phi_1 \\ \vdots \\ \phi_i \\ \vdots \\ \phi_n \end{bmatrix}^T \cdot \lim_{s \to \lambda_i} \begin{bmatrix} \frac{s-\lambda_i}{s-\lambda_1} & & & & \\ & \ddots & & & \\ & & \frac{s-\lambda_i}{s-\lambda_i} & & \\ & & & \ddots & \\ & & & & \frac{s-\lambda_i}{s-\lambda_n} \end{bmatrix} \cdot \begin{bmatrix} \psi_1 \\ \vdots \\ \psi_i \\ \vdots \\ \psi_n \end{bmatrix} \quad (29)$$
$$= \phi_i \psi_i$$

Combine (28) and (29):
$$\text{Res}_{\lambda_i}(sI-A)^{-1} = (\frac{\partial \lambda_i}{\partial A})^T \quad (30)$$

The partial derivatives of mode $\lambda_i$ with respect to the parameter $\rho$, along with the corresponding changes in the mode due to variations in the parameter, can be computed with (31) and (32). This enables a detailed analysis of how alterations in specific parameters influence the system's modal characteristics in the state space.
$$\frac{\partial \lambda_i}{\partial \rho} = \sum_{i=1}^{n}\sum_{j=1}^{n}(\frac{\partial \lambda_i}{\partial a_{ij}} \cdot \frac{\partial a_{ij}}{\partial \rho}) = -\text{tr}((\frac{\partial \lambda_i}{\partial A})^T \cdot \frac{\partial A}{\partial \rho})$$
$$= -\text{tr}(\text{Res}_{\lambda_i}(sI-A)^{-1} \cdot \frac{\partial A}{\partial \rho}) \quad (31)$$
$$\Delta \lambda_i = \frac{\partial \lambda_i}{\partial \rho} \cdot \Delta \rho \quad (32)$$

Comparing (12) and (31), the two modal sensitivity formulas have very similar expressions, highlighting the equivalence between MASS and MAI in terms of sensitivity analysis.

Note that MAI makes use of admittance matrix as the intermediate variable to calculate the modal sensitivity. The whole-system admittance matrix $Y$ inherently contains the position information, and more importantly, the partial derivative matrix $\partial Y / \partial y$ in (12) is sparse. The property makes modal analysis in the impedance model decoupled and narrowed down to a small group of admittance elements. In contrast, the state matrix $A$ is typically influenced by multiple factors, and computing $\partial A / \partial \rho$ of (31) is cumbersome. This key distinction between the two modal analysis methods underscores the adaptability of MAI for analyzing large and complex systems.

## IV. LAYER-WISE ENHANCEMENT OF MAI

As indicated in (18)-(20), [14] and [15] introduce a three-layer framework for root-cause investigations. Each layer offers varying levels of transparency based on the available prior knowledge. To improve the adaptability and accuracy across diverse scenarios, this section elaborates on a comprehensive discussion of layer-wise enhancement to MAI.

*A. First Layer Enhancement of MAI*

The first layer $\|s_{\lambda,y}\| \cdot \|y\|$ serves as an index that approximately describes the total participation degree of different elements when only apparatus impedance information is available. However, utilizing the Cauchy inequality for scaling might introduce relatively large errors to the index approximation. In addition, the first and second layers rely on the same prior information. Consequently, the index for all elements is updated using (33), incorporating the estimated damping and natural oscillation frequency derived from (19).

$$|\lambda| = \sqrt{\sigma_2^2 + \omega_2^2} \tag{33}$$

### B. Second Layer Enhancement of MAI

The analysis of series branches presented in [15] focuses solely on transmission lines, ignoring the potential errors introduced by the transformer branch. Consider an ideal transformer branch from bus $i$ to $j$ as shown in Fig. 2.

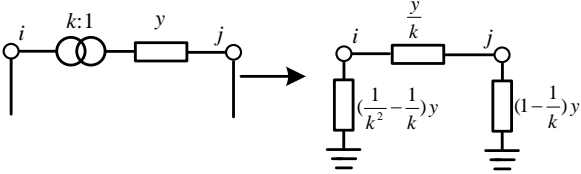

Fig. 2 Ideal transformer branch circuit.

The contribution of this ideal transformer branch to the whole-system admittance matrix is:

$$\begin{bmatrix} \Delta I_i \\ \Delta I_j \end{bmatrix} = y \begin{bmatrix} \dfrac{1}{k^2} & -\dfrac{1}{k} \\ -\dfrac{1}{k} & 1 \end{bmatrix} \begin{bmatrix} \Delta U_i \\ \Delta U_j \end{bmatrix} \tag{34}$$

Therefore, the transformer ratio $k$ will influence (12), and adjustments are necessary to rectify (13).

$$\frac{\partial \lambda}{\partial y} = -\mathrm{Res}_\lambda \left( \frac{Z_{ii}}{k^2} + Z_{jj} - \frac{Z_{ij}}{k} - \frac{Z_{ji}}{k} \right) \tag{35}$$

### C. Third Layer Enhancement of MAI

The third layer of MAI tracks instabilities to parameters using the parameter sensitivity factors defined in (16). In practice, $\partial y / \partial \rho$ is not easy to obtain because the impact of the perturbation injection amplitude on the errors of impedance measurement cannot be ignored [18].

An element with admittance $y$ is further divided into series branches and apparatus, like SGs or CIGs. For a series branch, the computation of $\partial y / \partial \rho$ can be simplified by splitting the impedance of a series branch into the inductive part and the resistive part, as shown in Fig. 3.

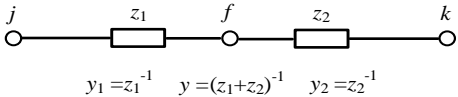

Fig. 3 Demonstration of line impedance split.

A virtual node $f$ splits a line into an inductive $z_1$ and a resistive part $z_2$:

$$z_1 = \begin{bmatrix} sL & -\omega_0 L \\ \omega_0 L & sL \end{bmatrix} = L \begin{bmatrix} s & -\omega_0 \\ \omega_0 & s \end{bmatrix}, s = \lambda \tag{36}$$

$$z_2 = \begin{bmatrix} R & 0 \\ 0 & R \end{bmatrix} = R \begin{bmatrix} 1 & 0 \\ 0 & 1 \end{bmatrix} \tag{37}$$

where $L$, $R$, and $\omega_0$ are the line inductance, resistance, and rated angular frequency of the system, respectively.

Subsequent to (20), the next objective is to compute $s_{\lambda,y}$ in (20), which involves calculating $Z_{if}$, $Z_{fj}$, $Z_{ff}$, and $Z_{jj}$ of the whole-system impedance matrix in (13). While $Z_{jj}$ remains unchanged, $Z_{if}$, $Z_{fj}$, and $Z_{ff}$ need to be recalculated using the system matrix post introduction of the virtual node $f$. This involves substantial computation and even necessitates recomputation for each new splitting. *Appendix A* outlines a novel procedure for directly solving $Z_{if}$, $Z_{fj}$, and $Z_{ff}$ from the original whole-system impedance matrix. Notably, the parameter sensitivity of the third layer has been facilitated by leveraging the information from the second layer, streamlining the analytical process significantly.

Regarding the apparatus, [19] highlights that the control link of a GFM converter can be viewed as circuit elements and represented by an impedance circuit model. This concept opens up the potential for transferring computations of $\partial y / \partial \rho$ of the apparatus to that of the series branches, as outlined earlier. However, further analyses are warranted to explore and validate this approach thoroughly.

## V. CASE STUDIES

### A. Overview of the Testing System and MAI Results

The modified IEEE 14-bus system with mixing of SGs and CIGs is considered, as shown in Fig. 4. The SGs are linked to buses 1, 2, 3, and 8, while the CIGs connected to bus 6 working in GFM control, and those connected to buses 11, 12, and 13 are in grid-following (GFL) control.

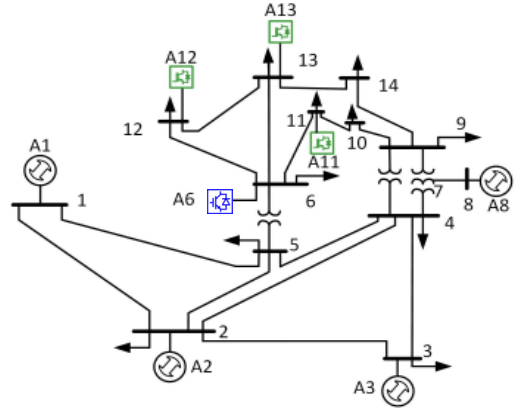

Fig. 4 Modification of IEEE 14-bus system.

The distribution of modes is depicted in Fig. 5. Consider the three modes marked by red crosses, which are proximate to the imaginary axis and represent different frequency bands. These three specific modes are utilized to demonstrate the effect of enhancement made to the first and second layers.

The results of the first and second layer of the original MAI for modes 1 and 2 are showcased through the headmap in Fig. 6. For each subgraph, the diagonal squares represent the participation degree of the power supply connected to the bus, and the non-diagonal squares denote the series branch existing between two distinct buses. Segments illustrated in white within the heatmap indicate the absence of an actual branch or power supply. This visualization aids in clearly understanding the influence and connectivity of various components within the system relative to the analyzed modes.

Fig. 6 (a) displays the results for the total participation degree of the first layer; the darker a square, the higher the total participation degree it represents. Thus, the GFM A6 exhibits the highest participation degree, indicating that this converter has the most substantial influence on mode 1. Fig. 6 (b) illustrates the changes in the real part of mode 1 resulting from an increase in the admittance of the corresponding element. A green coloration signifies a movement away from the imaginary axis, indicative of an increase in damping. Conversely, red indicates a movement closer to the imaginary axis, suggesting a decrease in damping, which can potentially reduce stability. Fig. 6 (c) depicts how the imaginary part of mode 1 changes when the admittance of an element is increased, providing insights into how adjustments in admittance affect the frequency characteristics of the mode.

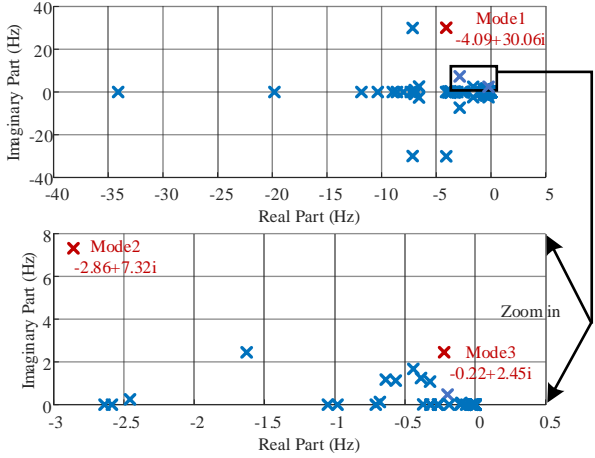

Fig. 5 Eigenvalue plot.

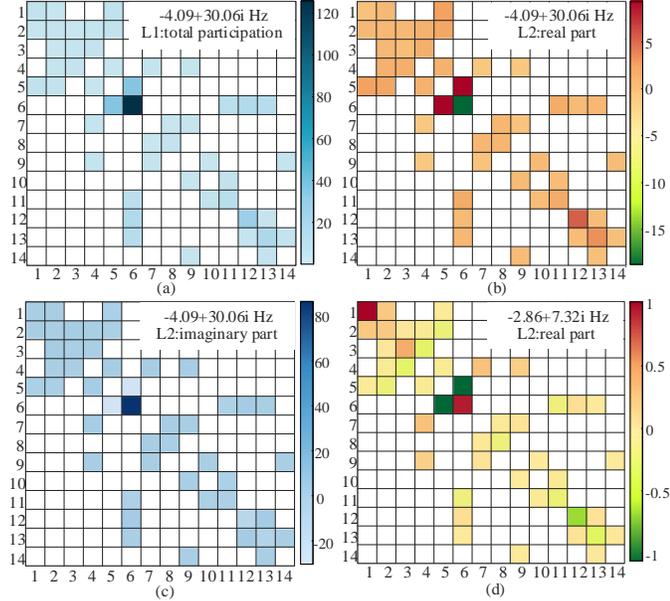

Fig. 6 Diagram of the first and second layer of MAI results for mode 1 and 2. (a) The first layer of mode 1. (b) Real part of the second layer of mode 1. (c) Imaginary part of the second layer of mode 1. (d) Real part of the second layer of mode 2.

It is essential to observe that when the admittance of branch 5-6 or the GFM converter A6 is changed, the direction of damping changes in modes 1 and 2 are opposite, as indicated by the differences between Fig. 6 (b) and (d). Adjusting system parameters to enhance the damping of mode 2, which is more susceptible to oscillations due to its proximity to the imaginary axis, will inadvertently reduce the damping in mode 1. This trade-off highlights the complexity of system tuning, where improvements in one aspect of system stability may compromise another.

### B. Evaluation of First Layer Enhancement

Regarding mode 3, Fig. 7 (a) reveals that GFM A6 continues to exhibit the highest participation degree for this mode. However, the use of the Cauchy inequality for scaling may lead to significant errors. The enhanced approximation of the participation degree, depicted in Fig. 7(b), points to SG A8 having a more substantial impact on mode 3, rather than GFM A6. This can be verified by observing the response of the angular frequency resulting from a small step change in load at the relevant buses in Fig. 8. The data shows that SG A8 exhibits a larger oscillation amplitude than GFM A6 at the oscillation frequency of 2.45 Hz, suggesting that SG A8 has a greater impact on mode 3. This supports the corrected participation degree shown in the enhanced layer 1 analysis from Fig. 7(b), demonstrating consistency with the simulation results.

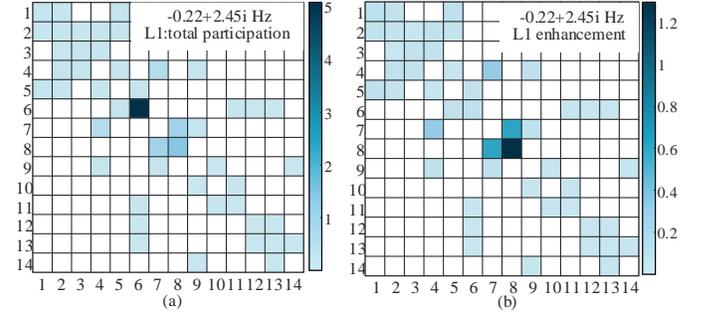

Fig. 7 Unenhanced and enhanced layer 1 of mode 3. (a) Unenhanced layer 1 of mode 3. (b) Enhanced layer 1 of mode 3.

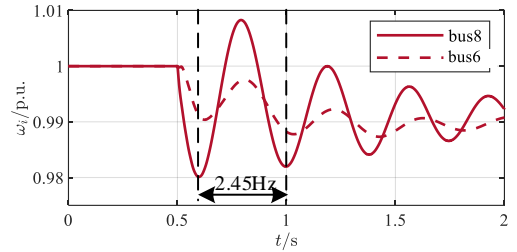

Fig. 8 The frequency change of SG A8 and GFM A6 after perturbation.

Table I presents the values computed for the first and second layers, both with and without the enhancement. This data clearly illustrates the differences in analysis outcomes, further emphasizing the potential discrepancies caused by scaling errors. To be specific, the maximum change in the eigenvalue estimated through the Cauchy inequality is indicated to be 5.104, whereas the real and imaginary parts of the second layer turn out to be only -0.047 and 0.004. This correction underscores the necessity to refine the methods employed in participation analysis to ensure accuracy.

TABLE I
LAYER 1, LAYER 2, AND ENHANCED LAYER 1 OF MODE 3

| Layer, $\|\lambda\|$ | GFM A6 | SG A8 |
| --- | --- | --- |
| Unenhanced Layer 1, $\|s_{\lambda,y}\| \cdot \|y\|$ | 5.104 | 1.357 |
| Layer 2 Real Part, $\sigma_2$ | -0.047 | 0.291 |
| Layer 2 Imaginary Part, $\omega_2$ | 0.004 | -1.262 |
| Enhanced Layer 1, $\sqrt{\sigma_2^2 + \omega_2^2}$ | 0.047 | 1.295 |

## C. Evaluation of Second Layer Enhancement

We use branch 5-6 of the modification of IEEE 14-bus system, which is a transformer branch, as an example to verify the enhancement of the second layer by considering the transformer ratio. Assume that the parameter $\rho$ of the branch has a 5% increase, the eigenvalue change of the mode is:

$$\Delta\lambda_{pr} = s_{\lambda,\rho} \cdot \rho \cdot 5\% \quad (38)$$

The error between the estimated and actual eigenvalue is:

$$\text{Error} = \frac{|\Delta\lambda_{pr} - \Delta\lambda_r|}{|\Delta\lambda_{pr}|} \quad (39)$$

where $\Delta\lambda_r$ denotes the actual eigenvalue change computed using the complete state-space model. Table II summarizes the findings for estimating changes in eigenvalues without incorporating transformer ratio enhancements using (13) and with the enhancements using (35) for mode 1 and mode 2. Analysis of the results in Table II reveals that by accounting for the transformer ratio adjustments, the error margin observed with the enhanced MAI is approximately 3%, compared to an error of about 15% when transformer ratio enhancements are not considered. This indicates that the error margin is roughly five times higher when the transformer ratio is omitted from the analysis.

TABLE II
ERROR ANALYSIS OF TRANSFORMER RATIO ENHANCEMENT ON MODAL ANALYSIS

| $\lambda,\rho$ | $\lambda_1, L_{56}$ | $\lambda_2, L_{56}$ |
|---|---|---|
| $\Delta\lambda_r$ | -0.0621 + 0.2059i | 0.0084 - 0.0871i |
| Unenhanced $\Delta\lambda_{pr}$ | -0.0741 + 0.2415i | 0.0084 - 0.1039i |
| Estimation Error | 14.854% | 16.146% |
| Enhanced $\Delta\lambda_{pr}$ | -0.0612 + 0.2129i | 0.0084 - 0.0901i |
| Estimation Error | 3.165% | 3.251% |

The line branch can be viewed as a special case of the transformer branch, with the transformer ratio set to 1. The findings in Table II are derived using a ratio value of 0.932. It is worth noting that the calculation accuracy is subject to the proximity of the ratio to 1; deviations from this value result in increased calculation errors. Accounting for the star-angle connection of the transformer introduces a complex ratio with a 30-degree angular shift, that will heighten the discrepancy in calculated results. Hence, considering the transformer ratio is imperative for ensuring accurate and reliable outcomes when applying MAI in practical scenarios.

## D. Evaluation of Third Layer Enhancement

In order to simulate the parameter change scenario of the third layer, the inductance of branch 5-6 has uniformly varied a total of 7 times, with each modification reducing the inductance to 0.8 times its preceding value. The estimated modal changes are depicted by stars, while the actual changes are shown by crosses in Fig. 9. Prior to the parameter change, the eigenvalue of the system was recorded at -1.16 + 4.66i. After the series of inductance modifications, the eigenvalue shifted to -1.34 + 6.22i. The alignment between the estimated and actual data underscores the effectiveness of the proposed approach in accurately estimating modal shifts due to parameter variations, thereby affirming the robustness and reliability of the MAI enhancement.

In addition, Fig. 10 depicts the time-domain response curves of the angular frequency associated with GFM A6 both before and after the inductance variation following a minor load alteration at bus 6. The observations reveal an increment in the oscillation frequency and a corresponding decline in the peak oscillation magnitude post the inductance modification. This trend indicates a reduction in the real part and an elevation in the imaginary part of the eigenvalue pertaining to the relevant mode. Furthermore, an approximate estimation of the frequency suggests a shift from around 4.66Hz pre-parameter change to 6.22Hz post-adjustment. Remarkably, these estimations align closely with the values computed utilizing the enhanced MAI technique illustrated in Fig. 9.

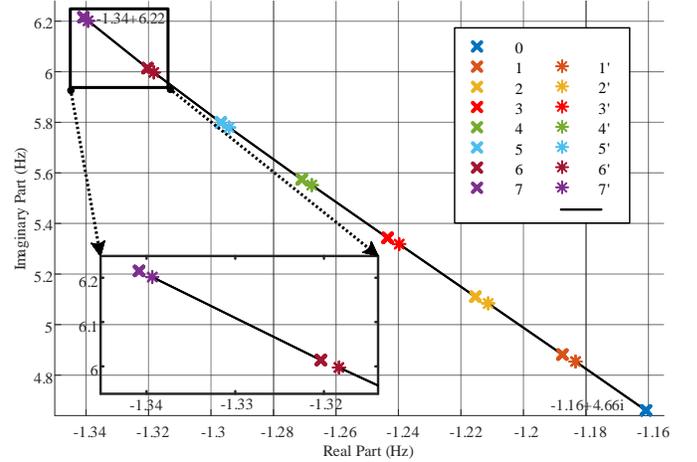

Fig. 9 Mode before and after branch parameter adjustment.

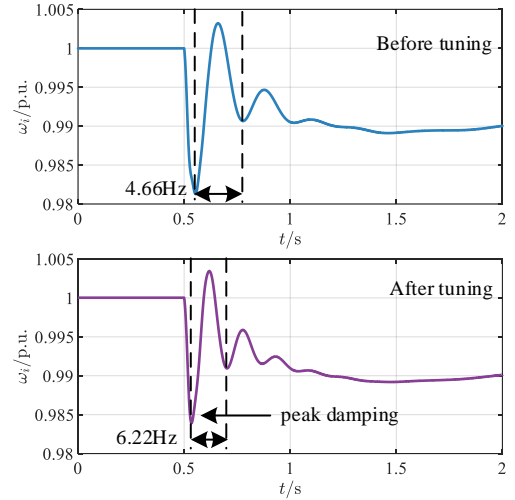

Fig. 10 Frequency diagram of SG A8 with 1 p.u. increase in bus 6 active load at 0.5 sec. before and after branch circuit parameter adjustment.

## VI. CONCLUSION

MAI plays a pivotal role in analyzing the involvement of elements across various modes within power systems with high CIG penetration. This paper establishes, for the first time, the equivalence between MASS and MAI in terms of transfer function, eigenvalues, and sensitivity. This pioneering comparison establishes a theoretical basis for identifying elements with the most significant participation across system modes using solely impedance details. In addition, layer-wise enhancement of the original MAI method is introduced and verified by simulations on a modified IEEE 14-bus system. Future work will focus on considering the influence of various control loops of the converter in the impedance modeling and MAI framework.

## APPENDIX A

As shown in Fig. 3, after adding a new node $f$, $Z_{if}$, $Z_{fi}$, and $Z_{ff}$ can be calculated from the original whole-system impedance matrix.

Consider injecting current $I_i$ to any node $i$ of the system, then the voltage of node $f$ is

$$U_f = Z_{fi}I_i = U_j - z_1 I_{jk} = U_j - z_1 y(U_j - U_k) \\ = (Z_{ji} - z_1 y(Z_{ji} - Z_{ki}))I_i \tag{A1}$$

Therefore, $Z_{fi}$ is:

$$Z_{fi} = Z_{ji} - z_1 y(Z_{ji} - Z_{ki}) \tag{A2}$$

Replacing $i$ with $j$ yields $Z_{fj}$ as:

$$Z_{fj} = Z_{jj} - z_1 y(Z_{jj} - Z_{kj}) \tag{A3}$$

Considering the injection of current $I_f$ into node $f$, the voltage of node $i$ is:

$$U_i = Z_{if} I_f \tag{A4}$$

Assemble the node conductance matrix for nodes $j$, $k$, and $f$ and their branch columns:

$$\begin{bmatrix} 0 \\ 0 \\ I_f \end{bmatrix} = \begin{bmatrix} z_1^{-1} & 0 & -z_1^{-1} \\ 0 & z_2^{-1} & -z_2^{-1} \\ -z_1^{-1} & -z_2^{-1} & z_1^{-1} + z_2^{-1} \end{bmatrix} \begin{bmatrix} U_j \\ U_k \\ U_f \end{bmatrix} \tag{A5}$$

Implement matrix shrinkage to remove node $f$ in (A5), and the injected current $I_f$ at node $f$ is transferred to nodes $j$ and $k$ as follows:

$$\begin{bmatrix} I_j \\ I_k \end{bmatrix} = \begin{bmatrix} z_1^{-1} \\ z_2^{-1} \end{bmatrix}(z_1^{-1} + z_2^{-1})^{-1} I_f \tag{A6}$$

Then the voltage of node $i$ can be expressed as:

$$U_i = Z_{ij}I_j + Z_{ik}I_k = (Z_{ij}z_1^{-1} + Z_{ik}z_2^{-1})(z_1^{-1} + z_2^{-1})^{-1} I_f \tag{A7}$$

Comparing (A7) with (A4), $Z_{if}$ is as follows:

$$Z_{if} = (Z_{ij}z_1^{-1} + Z_{ik}z_2^{-1})(z_1^{-1} + z_2^{-1})^{-1} \tag{A8}$$

Using the same method we can simply get $Z_{jf}$ as well.

$$Z_{jf} = (Z_{jj}z_1^{-1} + Z_{jk}z_2^{-1})(z_1^{-1} + z_2^{-1})^{-1} \tag{A9}$$

Considering current $I_f$ injected into the new node $f$, the voltages of node $f$, $j$, and $k$ are:

$$U_f = Z_{ff} I_f, U_j = Z_{jf} I_f, U_k = Z_{kf} I_f \tag{A10}$$

Writing the Kirchhoff's Current Law (KCL) equation for the column of node $f$ yields:

$$z_1^{-1}(U_f - U_j) + z_2^{-1}(U_f - U_k) = I_f \tag{A11}$$

Substituting (A10) into (A11) gives:

$$z_1^{-1}(Z_{ff} - Z_{jf}) + z_2^{-1}(Z_{ff} - Z_{kf}) = \boldsymbol{I} \tag{A12}$$

where $\boldsymbol{I}$ represent the unit matrix.

Finally, $Z_{ff}$ is determined utilizing (A12):

$$Z_{ff} = (z_1^{-1} + z_2^{-1})^{-1}(\boldsymbol{I} + z_1^{-1}Z_{jf} + z_2^{-1}Z_{kf}) \tag{A13}$$

## REFERENCES

[1] Y. Gu and T. C. Green, "Power system stability with a high penetration of inverter-based resources," *Proc. IEEE Proc. IRE*, vol. 111, no. 7, pp. 832-853, July 2023.
[2] Y. Cheng et al., "Real-world subsynchronous oscillation events in power grids with high penetrations of inverter-based resources," *IEEE. Trans. Power Syst.*, vol. 38, no. 1, pp. 316-330, Jan. 2023.
[3] H. Cui et al., "Disturbance propagation in power grids with high converter penetration," *Proc. IEEE Proc. IRE*, vol. 111, no. 7, pp. 873-890, July 2023.
[4] Y. Li, Y. Gu, and T. C. Green, "Revisiting grid-forming and grid-following inverters: A duality theory," *IEEE Trans. Power Syst.*, vol. 37, no. 6, pp. 4541-4554, Nov. 2022.
[5] W. Cao, Y. Ma, L. Yang, F. Wang and L. M. Tolbert, "D–Q impedance based stability analysis and parameter design of three-phase inverter-based AC power systems," *IEEE Trans. Ind. Electron.*, vol. 64, no. 7, pp. 6017-6028, July 2017.
[6] M. Lu, Y. Yang, B. Johnson, and F. Blaabjerg, "An interaction-admittance model for multi-inverter grid-connected systems," *IEEE Trans. Power Electron.*, vol. 34, no. 8, pp. 7542-7557, Aug. 2019.
[7] L. Orellana, L. Sainz, E. Prieto-Araujo and O. Gomis-Bellmunt, "Stability assessment for multi-infeed grid-connected VSCs modeled in the admittance matrix form," *IEEE Trans. Circuits Syst. I, Reg. Papers*, vol. 68, no. 9, pp. 3758-3771, Sept. 2021.
[8] Y. Gu, Y. Li, Y. Zhu, and T. C. Green, "Impedance-based whole-system modeling for a composite grid via embedding of frame dynamics," *IEEE Trans. Power Syst.*, vol. 36, no. 1, pp. 336-345, Jan. 2021.
[9] P. Kundur, *Power System Stability and Control*. New York, NY, USA: McGraw-Hill, vol. 7, 1994.
[10] Wilsun Xu, Zhenyu Huang, Yu Cui, and Haizhen Wang, "Harmonic resonance mode analysis," *IEEE Trans. Power Del.*, vol. 20, no. 2, pp. 1182-1190, April 2005.
[11] Z. Huang, Y. Cui, and W. Xu, "Application of modal sensitivity for power system harmonic resonance analysis," *IEEE Trans. Power Syst.*, vol. 22, no. 1, pp. 222-231, Feb. 2007.
[12] Y. Wang, X. Wang, F. Blaabjerg and Z. Chen, "Harmonic instability assessment using state-space modeling and participation analysis in inverter-fed power systems," *IEEE Trans. Ind. Electron.*, vol. 64, no. 1, pp. 806-816, Jan. 2017.
[13] Y. Zhan, X. Xie, H. Liu, H. Liu, and Y. Li, "Frequency-domain modal analysis of the oscillatory stability of power systems with high-penetration renewables," *IEEE Trans. Sustain. Energy*, vol. 10, no. 3, pp. 1534-1543, July 2019.
[14] Y. Zhu, Y. Gu, Y. Li, and T. C. Green, "Participation analysis in impedance models: The grey-box approach for power system stability," *IEEE Trans. Power Syst.*, vol. 37, no. 1, pp. 343-353, Jan. 2022.
[15] Y. Zhu, Y. Gu, Y. Li, and T. C. Green, "Impedance-based root-cause analysis: Comparative study of impedance models and calculation of eigenvalue sensitivity," *IEEE Trans. Power Syst.*, vol. 38, no. 2, pp. 1642-1654, March 2023.
[16] Q. Zheng, F. Gao, Y. Li, Y. Zhu, and Y. Gu, "Equivalence of impedance participation analysis methods for hybrid AC/DC power systems," *IEEE Trans. Power Syst.*, vol. 39, no. 2, pp. 3560-3574, March 2024.
[17] B. Gustavsen and A. Semlyen, "Rational approximation of frequency domain responses by vector fitting," *IEEE Trans. Power Del.*, vol. 14, no. 3, pp. 1052-1061, July 1999.
[18] Y. Zhu, Y. Zhang, and T. C. Green, "Injection amplitude guidance for impedance measurement in power systems," *IEEE Trans. Power Electron.*, vol. 38, no. 6, pp. 6929-6933, June 2023.
[19] Y. Li, Y. Gu, Y. Zhu, A. Junyent-Ferré, X. Xiang and T. C. Green, "Impedance circuit model of grid-forming inverter: Visualizing control algorithms as circuit elements," *IEEE Trans. Power Electron.*, vol. 36, no. 3, pp. 3377-3395, March 2021.